\documentclass{emulateapj}
\usepackage{apjfonts}

\begin{document} 
%%%%%%%%%%%%%%%%%%%%%%%%%%%%%%%%%%%%%%%%%
%%%%%%%%%%%%%%%%%%%%%%%%%%%%%%%%%%%%%%%%%

\title{Superbursts from Strange Stars}

\author{Dany Page\altaffilmark{1} and Andrew Cumming\altaffilmark{2}}
\altaffiltext{1}{Instituto de Astronom\'ia, 
                 Universidad Nacional Aut\'onoma de M\'exico, 
                 04510 Mexico D.F., Mexico.}
\altaffiltext{2}{Physics Department, McGill University, 
                 3600 rue University, Montreal, QC, H3A 2T8, Canada.}

%%%%%%%%%%%%%%%%%%%%%%%%%%%%%%%%%%%%%%%%%
\begin{abstract} 
Recent models of carbon ignition on accreting neutron stars predict superburst ignition depths that are an order of magnitude larger than observed. We explore a possible solution to this problem, that the compact stars in low mass X-ray binaries that have shown superbursts are in fact strange stars with a crust of normal matter. We calculate the properties of superbursts on strange stars, and the resulting constraints on the properties of strange quark matter. We show that the observed ignition conditions exclude fast neutrino emission in the quark core, for example by the direct Urca process, which implies that strange quark matter at stellar densities should be in a color superconducting state. For slow neutrino emission in the quark matter core, we find that reproducing superburst properties requires a definite relation between three poorly constrained properties of strange quark matter: its thermal conductivity ($K$), its slow neutrino emissivity ($\epsilon_\nu \simeq Q_\nu \times T_9^8$) and the energy released by converting a nucleon into strange quark matter ($Q_\mathrm{SQM}$).
\end{abstract} 
%%%%%%%%%%%%%%%%%%%%%%%%%%%%%%%%%%%%%%%%%


\maketitle 

%%%%%%%%%%%%%%%%%%%%%%%%%%%%%%%%%%%%%%%%%
\section{Introduction}
         \label{Sec:Intro}
%%%%%%%%%%%%%%%%%%%%%%%%%%%%%%%%%%%%%%%%%

Strange stars are compact stars that are almost entirely made of deconfined quark matter, long ago suggested as an alternative to neutron stars following the proposal that strange quark matter (SQM) is the ground state of hadronic matter (\citealt{W84}; see also \citealt{I70}) (For a review see \citealt{W05}). 
Strange quark matter with a density of $\sim 5\times 10^{14}$ g~cm$^{-3}$ could exist up to the surface of a strange star, in which case the star would look, and evolve, very different from a neutron star \citep{PU02}. However, a strange star can also have a crust of normal baryonic matter \citep{AFO86}, in which case its surface emission properties would be the same as those of a neutron star. The bulk properties of both kinds of stars, such as radius and moment of inertia, are also similar in the observed mass range $1\!< \! M/M_\odot \! <2$, and so in general it is difficult to distinguish between strange stars and neutron stars \citep{HZS86}. To date, there is no real evidence against or in favor of the strange star hypothesis \citep{PLSP03}, mostly due to the fact that many properties of quark matter are
poorly understood, allowing great freedom to vary parameters of the models to reproduce observed properties. An observable, and observed, characteristic of compact stars that would allow us to distinguish between these two very different kinds of objects has not yet been identified.

Superbursts are rare, extremely energetic, and long duration Type I X-ray bursts observed from low mass X-ray binaries (LMXBs) \citep{K04}. They have been modeled as being due to unstable thermonuclear burning of carbon on the surface of an accreting neutron star \citep{CB01, SB02}. The accreted hydrogen and helium burns shortly after its arrival on the stellar surface, producing a mixture of heavy elements and carbon ($X_C\sim 10$\% carbon by mass) that accumulates and later ignites. \cite{B04} and \cite{CN05} showed that the ignition conditions for superbursts are sensitive to the thermal properties of the stellar interior, in particular the composition of the crust, and the neutrino emission from the core. This opens up a new way to probe the interiors of the compact stars accreting in LMXBs.

However, calculations so far suggest that carbon ignition models, on a neutron star, have difficulty in achieving the low ignition column depths inferred from observations \citep{CB01, B04, CN05, Cetal05}. In particular, \cite{Cetal05} carefully constrained the ignition depth by fitting models of the cooling of the burning layer following a superburst \citep{CM04} to observed superburst lightcurves. They then showed that the inferred ignition column depths of $y=(0.5$--$3)\times 10^{12}\ {\rm g\ cm^{-2}}$ were an order of magnitude smaller than carbon ignition models predict. In particular, they showed that neutrino cooling in the crust from Cooper pairing of the superfluid neutrons \citep{FRS76} 
limits the crust temperature to $\lesssim 5\times 10^8\ {\rm K}$, making it impossible to achieve ignition at the inferred depths.

In this paper, we investigate a possible solution to this puzzle, which is that the compact stars in LMXBs that have shown superbursts are in fact strange stars with a crust of normal matter. These stars do not have an inner crust, and so neutron Cooper pair neutrino emission is absent; in addition, the energy release at the base of the crust from conversion of accreted nucleons into strange quark matter is potentially much greater than the energy release in neutron star crusts from pycnonuclear reactions and electron captures. We therefore expect the accreted crust of a strange star to be hotter than for a neutron star, making carbon ignition at column depths $\lesssim 10^{12}\ {\rm g\ cm^{-2}}$ possible. 
We first describe our strange star models in \S~\ref{Sec:Physics}, and then present calculations of carbon ignition in \S~\ref{Sec:Superbursts2}. We conclude in \S~\ref{Sec:Conclusion}.

%%%%%%%%%%%%%%%%%%%%%%%%%%%%%%%%%%%%%%%%%
\section{Strange Star Models}
         \label{Sec:Physics}
%%%%%%%%%%%%%%%%%%%%%%%%%%%%%%%%%%%%%%%%%

We first discuss our models of strange stars with a thin baryonic crust, and the physical parameters relevant for carbon ignition. The accumulating carbon layer is heated by the energy released at the base of the crust from conversion of nucleons into strange quark matter. Most of this energy is conducted into the core and escapes as neutrinos, but a small amount flows outwards through the crust and sets the ignition conditions. The important ingredients are (1) the crust thickness, (2) the energy released per nucleon by conversion into strange quark matter, $Q_\mathrm{SQM}$, (3) the neutrino emissivity $\epsilon_\nu$ and (4) thermal conductivity $K$ of quark matter. Given our poor understanding of quark matter these four ingredients are essentially free parameters and we will treat them as such, calculating the values required to match observed superburst properties.

\begin{figure}
\epsscale{1.1}
\plotone{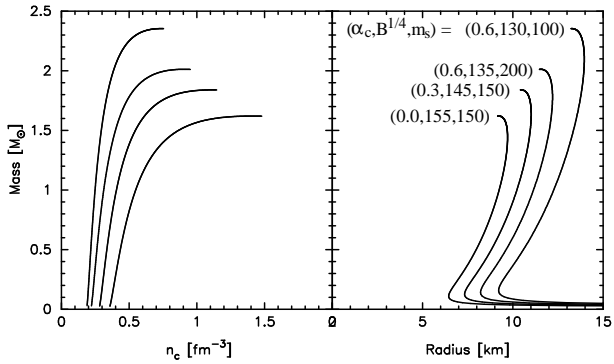}
\caption{Masses of strange stars vs central density (left) and radius (right) for four sets of values of the coupling constant $\alpha_c$, bag constant $B^{1/4}$ (in MeV), and strange quark mass $m_s$ (in MeV/c$^2$) covering a broad range of possible values.}
\label{Fig1}
\end{figure}

To obtain estimates of the possible range of values for the crust thickness and $Q_\mathrm{SQM}$ we consider the strange quark matter models of \cite{FJ84}. These MIT bag inspired models use three parameters, the bag constant $B$ which is the energy density cost for having deconfined quark matter, the strange quark mass $m_s$, and the color interaction coupling constant $\alpha_c$. For a broad range of values of these parameters strange quark matter is self bound. Figure \ref{Fig1} shows examples of mass-radius and mass-central density relationships for a broad range of parameters.
For definiteness, in our numerical models we choose $\alpha_c = 0.3$, $B$=(140 MeV)$^4$, and $M=1.4\ M_\odot$.

%------------------------------------------------------------------------------------------------
%------------------------------------------------------------------------------------------------

\begin{figure}
\epsscale{0.85}
\plotone{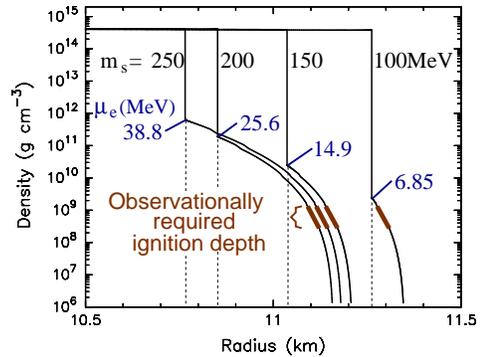}
\caption{Density profiles of the outer layers of four strange stars with crusts. The models have different values of $m_s$ but the same values of $\alpha_c=0.3$ and $B^{1/4}=140$ MeV. The electron chemical potential at the surface of the quark matter is indicated, as well as the inferred range of carbon ignition density for superbursts. The crust EOS is from \cite{HZ90}.}
\label{Fig2} 
\end{figure}

The presence of a crust of baryonic matter is possible because of the positive charge of strange quark matter which hence must contain electrons\footnote{We do not consider models having negatively charged quark matter which would contain positrons instead of electrons.}. The electrons are bound to the quark matter by the long range Colomb force, and leak out of the quark matter creating a dipolar layer of thickness a few thousand fms. The outward pointing electric field repels ions, electrostatically suspending the crust above the quark matter \citep{AFO86}. The existence of the supporting electric field requires that the electron chemical potential $\mu_e$ in the crust be lower than in the quark matter. This sets the thickness of the crust. Accretion drives continuous energy release, because as the density at the base of the crust increases, the value of $\mu_e$ there grows and approaches its value at the surface of the quark matter resulting in a reduction, or even inversion, of the electric field:  ions plunge into the quark matter, releasing an energy $Q_\mathrm{SQM}$ per baryon. The thickest possible crust, which occurs when $\mu_e$ in quark matter is larger than $\sim$ 30 MeV, reaches neutron drip at its base: the dripped neutrons fall directly into the quark matter, also producing an energy $Q_\mathrm{SQM}$. Within the \cite{FJ84} model, for fixed $B$ and $\alpha_c$, the value of $\mu_e$ at the surface of the quark matter core is determined by the values of $m_s$, a larger $m_s$ resulting in a larger $\mu_e$. Figure \ref{Fig2} gives the density profile of the outer layers of the four strange star models we will use. The thickness of the nuclear crust ranges from $\approx 100\ {\rm m}$ to a maximum extent, at neutron drip, of $\approx 400\ {\rm m}$.

The energy released per nucleon by converting it into strange quark matter, $Q_\mathrm{SQM}$,  is our second free parameter. 
Within the \cite{FJ84} model it can take values from a few MeV up to almost 100 MeV,
depending on the precise values of the model parameters ($B$, $\alpha_c$, $m_s$).
For our numerical simulations we vary $Q_\mathrm{SQM}$ over a broad range\footnote{
Strictly speaking, varying $Q_\mathrm{SQM}$ means varying the EOS while we use only four cases of EOS and vary $Q_\mathrm{SQM}$ for each.
However our results are only weakly dependent on the mass and size of the star and adjusting the EOS for each value of $Q_\mathrm{SQM}$ would not
affect them. However, within the \cite{FJ84} models large values of $m_s$ result in a reduction of the maximum possible value of $Q_\mathrm{SQM}$ and also an increase in $\mu_e$, i.e., thicker crusts: cases which allow $\mu_e >$ 30 MeV, and hence a maximally thick crust, limit $Q_\mathrm{SQM}$ to be less than 40 MeV.
Considering that this is only an approximate model we choose to explore the broadest possible range of values of $Q_\mathrm{SQM}$ but one must keep in mind this possible restriction of $Q_\mathrm{SQM}$ in analyzing our final results.}, from $<1$ MeV to $>100$ MeV.

%------------------------------------------------------------------------------------------------
%------------------------------------------------------------------------------------------------

The neutrino emissivity of quark matter can span the broadest possible range.
If quarks form a normal Fermi liquid, the direct Urca process,
$u + e^- \rightarrow d + \nu_e$ with $d \rightarrow u + e^- + \overline{\nu}_e$
(and the same process with an $s$ replacing the $d$), has an enormous emissivity of the
order of $\epsilon_\nu^\mathrm{DUrca} \sim 10^{25} \; T_9^6$ erg cm$^{-3}$ s$^{-1}$,
where $T_9$ is the temperature $T$ in units of $10^9$ K.
However, in the very likely situation that some flavors/colors of quarks form a color superconductor \citep{A01}, the neutrino emission may be determined by bremsstrahlung from unpaired quarks, or from electrons (e.g., \citealt{PGW05}). We write these slow neutrino emission processes as
\begin{eqnarray}
\epsilon_\nu \simeq Q_\nu \times T_9^8 \; \mathrm{erg \; cm^{-3} \; s^{-1}}
\label{Eq:neutrinos}
\end{eqnarray}
and consider $Q_\nu$ as our third free parameter with values from about $10^{18}$ up to an optimistic maximum of $\sim 10^{22}$.

Finally, the thermal conductivity $K$ of quark matter is our last free parameter.
We will consider values from $K=10^{19}$ to $10^{22}$ erg cm$^{-1}$ K$^{-1}$,
which cover the whole range of possible values (e.g., \citealt{PU02}).

%%%%%%%%%%%%%%%%%%%%%%%%%%%%%%%%%%%%%%%%%
\section{Carbon Ignition in Strange Stars}
         \label{Sec:Superbursts2}
%%%%%%%%%%%%%%%%%%%%%%%%%%%%%%%%%%%%%%%%%

%------------------------------------------------------------------------------------------------------------
\begin{figure}
\epsscale{1.1}
\plotone{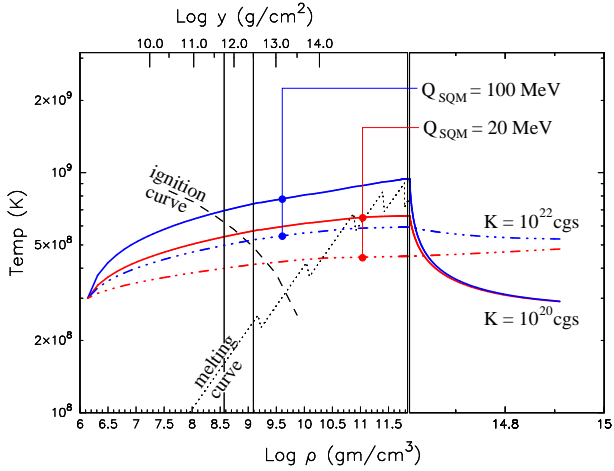}
\caption{Temperature profiles for strange stars with a thick crust reaching the neutron drip density, accreting at $\dot{m} = 0.3 \dot{m}_\mathrm{Edd}$. We take a slow neutrino emissivity (eq.~\ref{Eq:neutrinos}) with $Q_\nu = 10^{21}$, but different values of $K$ and $Q_\mathrm{SQM}$ as indicated. The carbon ignition curve (with $X_C=0.2$) is shown as a dashed line. The two vertical lines indicate the observed range of ignition depths.
The dotted curve shows the melting line. Note that most of the crust is liquid (in fact, we find that thinner crusts can be completely liquid).
 \label{Fig3}}
\end{figure}
%------------------------------------------------------------------------------------------------------------

We now present our numerical calculations of the thermal state of an accreting strange star, and investigate the range of values of parameters that result in successful superbursts, i.e., carbon ignition at column depths of $y \approx 10^{12}$ g cm$^{-2}$.
We assume a constant accretion rate $\dot{m}$ (in nucleons cm$^{-2}$ s$^{-1}$)
and follow the time evolution of the star until it has reached a stationary state.
Typical values of $\dot{m}$ are about 0.1--0.3 $\dot{m}_\mathrm{Edd}$
(in terms of the Eddington rate $4\pi R^2 m_u \dot{m}_\mathrm{Edd} \simeq 10^{18}$ g s$^{-1}$, $R$ being the star's radius and $m_u$ the atomic mass unit).
Heat is injected at the top of the strange quark matter core at a rate given by $H = Q_\mathrm{SQM} \times 4\pi R^2 \dot{m}$.
Our code solves the heat transport and energy balance equations in full general relativistic form with spherical symmetry.
%The structure of the star is obtained by solving the Tolman-Oppenheimer-Volkoff equation of hydrostatic equilibrium for the given strange quark matter and crust EOSs (Fig.~\ref{Fig2}).

A first and firm result of our simulations is that it is impossible for the crust to reach temperatures close to the required $\approx 6 \times 10^8$ K if a direct Urca process operates in the core, even if suppressed down to $\epsilon_\nu \sim 10^{22} T_9^6$ erg cm$^{-3}$ s$^{-1}$. Therefore, the neutrino emissivity of strange quark matter must be restricted to the slow processes of equation (\ref{Eq:neutrinos}).
Illustrative examples of temperature profiles are shown in Figure \ref{Fig3}, assuming $Q_\nu=10^{21}\ {\rm erg\ cm^{-3}\ s^{-1}}$, but different values of $K$ and $Q_{SQM}$. The model with $K=10^{20}$ cgs and $Q_\mathrm{SQM}=100$ MeV is too hot, and ignites at a column depth smaller than observed, whereas the model with 
$K=10^{22}$ cgs and $Q_\mathrm{SQM}=20$ MeV is too cold, and ignites at a larger column than observed. However, we obtain ignition at the appropriate depth for either $K=10^{22}$ cgs and $Q_\mathrm{SQM}=100$ MeV, or $K=10^{20}$ cgs and $Q_\mathrm{SQM}=20$ MeV.

We can understand the temperature profiles in Figure \ref{Fig3} with a simple analytic model. Because the crust is approximately isothermal below the ignition depth, the temperature at the crust-core interface is a reasonable estimate of the ignition temperature. In addition, most of the heat released at this interface flows into the core (\citealt{Cetal05} showed that an outwards flux of only $Q_b\approx 0.25\ {\rm MeV\ per\ nucleon}\ (\dot m/0.3 \dot  m_{\rm Edd})^{-1}\ll Q_{SQM}$ is needed to achieve ignition at $y\approx 10^{12}\ {\rm g\ cm^{-2}}$). When the thermal conductivity is large, this heat is rapidly distributed over the core, which is then isothermal at a temperature given by balancing the heating with the energy loss due to neutrinos: $4\pi R^2\dot m Q_{SQM}=(4\pi R^3/3)Q_\nu T_9^8$. We therefore find
that $T_9=(3\dot m Q_\mathrm{SQM}/R Q_\nu)^{1/ 8}$, or
\begin{eqnarray}
\label{eq:iso}
T_9=0.54\ Q_{\nu, 21}^{-1/8}\left({Q_{SQM}\over 100\ {\rm MeV}}\right)^{1/8}
\left({\dot m\over 0.3\dot m_{\rm Edd}}\right)^{1/8}
R_6^{-1/8},
\end{eqnarray}
where $Q_{\nu,21}=Q_{\nu}/10^{21}\ {\rm erg\ cm^{-3}\ s^{-1}}$ and $R_6=R/10\ {\rm km}$. This estimate agrees well with Figure \ref{Fig3} for models with an isothermal core ($K=10^{22}\ {\rm cgs}$). 

However, when the thermal conductivity is low, the core is no longer isothermal, and we must estimate the temperature profile in the core. To do this, we use an approximate plane parallel model of the core. The temperature profile is given by $\rho dF/dy=-\epsilon_\nu$, where $F=-\rho K dT/dy$ is the inwards directed heat flux, $\rho$ being the density. Assuming $\rho$, $Q_\nu$ and $K$ are constant in the core gives $d^2 T/dy^2=Q_\nu T_9^8/ \rho^2 K$. The solution is $T\propto (y+b)^{-2/7}$, where $b$ is a constant determined by the requirement that all of the heat flux entering the core is emitted as neutrinos: $\dot mQ_\mathrm{SQM}=\int \epsilon_\nu dy$. Assuming the surface contribution dominates the integral, we find the interesting result $(\dot mQ_\mathrm{SQM})^2=(2/9)(RQ_\nu T^8)(KT/R)$; the heat flux entering the core is the geometric mean of the neutrino emission at the core surface and the heat flux transported deep into the core. The temperature at the core surface is
\begin{eqnarray}
\label{eq:noniso}
T_9=0.87\ Q_{\nu,21}^{-1/9}K_{20}^{-1/9}
\left({Q_{SQM}\over 100\ {\rm MeV}}\right)^{2/9}
\left({\dot m\over 0.3\ \dot m_{\rm Edd}}\right)^{2/9},
\end{eqnarray}
where $K_{20}=K/10^{20}\ {\rm cgs}$, which agrees well with the non-isothermal core profiles in Figure \ref{Fig3}.
Comparing equations (\ref{eq:iso}) and (\ref{eq:noniso}) gives the thermal conductivity of the core at which the temperature profile becomes isothermal as $K_{\rm crit}=4.3\times 10^{21}\ {\rm cgs}\ Q_{\nu,21}^{1/8}(Q_{SQM}/100\ {\rm MeV})^{7/8}(\dot m/0.3\ \dot m_{\rm Edd})^{7/8}$.

%------------------------------------------------------------------------------------------------------------
\begin{figure}
\epsscale{0.95}
\plotone{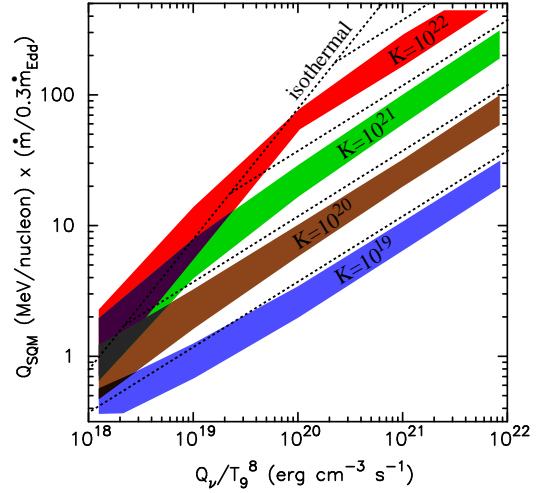}
\caption{The energy release $Q_\mathrm{SQM}$ required for carbon ignition at $10^{12}\ {\rm g\ cm^{-2}}$ as a function of $Q_\nu$ and $K$. The thickness of each shaded stripe corresponds to the range of crust thickness in our models (Fig.~\ref{Fig2}; a thicker crust requires a larger $Q_{SQM}$). The criterion for ignition is $T=6\times 10^8$ K at $y=10^{12}$ g cm$^{-2}$. To account for the allowed range of $y$ shown by the vertical solid lines in Figure \ref{Fig3}, the upper limit of each strip can be increased by a factor two (corresponding to $y \simeq 0.5 \times 10^{12}$ g cm$^{-2}$), or the lower limit decreased by a factor two ($y \simeq 3 \times 10^{12}$ g cm$^{-2}$). The dashed lines show the analytic estimates for the same values of $K$.}
\label{Fig4}
\end{figure}
%------------------------------------------------------------------------------------------------------------

Figure \ref{Fig4} shows the combinations of $Q_\mathrm{SQM}$, $Q_\nu$, and $K$ that are needed for ignition at $y=10^{12}\ {\rm g\ cm^{-2}}$. We find the surprising result that even values of $Q_\mathrm{SQM}$ as large as 100 MeV (compared to the $\approx 1.5$ MeV released by pycnonuclear reactions in the neutron star case) can give the required crust temperature when $K$ and $Q_\nu$ are sufficiently large. The shape of the curves is determined by the fact that, at a fixed value of $Q_\nu$, as the conductivity $K$ is increased eventually the core becomes isothermal, and the solution then becomes independent of $K$. For an isothermal core, fixing the temperature at the boundary to be $T_9=0.7$ (which gives ignition at the right depth; Fig.~\ref{Fig3}) in equation (\ref{eq:iso}) gives $Q_\mathrm{SQM,iso}=80\ {\rm MeV}\ Q_{\nu,20}R_6(\dot m/0.3\ \dot m_{\rm Edd})^{-1}$. Similarly, setting $T_9=0.7$ in equation (\ref{eq:noniso}) for the non-isothermal case gives $Q_\mathrm{SQM,noniso}=38\ {\rm MeV}\ Q_{\nu,21}^{1/2}K_{20}^{1/2}(\dot m/0.3\ \dot m_{\rm Edd})^{-1}$. These analytic estimates are plotted as dotted lines in Figure \ref{Fig4}, and nicely reproduce the numerical results.

%%%%%%%%%%%%%%%%%%%%%%%%%%%%%%%%%%%%%%%%%
\section{Discussion and Conclusions}
         \label{Sec:Conclusion}
%%%%%%%%%%%%%%%%%%%%%%%%%%%%%%%%%%%%%%%%%

In this paper, we have explored the scenario that superbursts originate on strange stars with a thin baryonic crust.  Given our poor understanding of quark matter, it is perhaps not surprising that we have been able to find many choices of parameters for accreting strange stars that reproduce the observed properties of superbursts. Nevertheless, the models provide significant constraints on the properties of strange quark matter. In particular, we cannot reproduce the observed ignition conditions if the neutrino emission in the quark matter core is by a fast process such as direct Urca. The fact that direct processes must be suppressed implies that strange quark matter at stellar densities should be a color superconductor \citep{A01,PGW05}. For slow neutrino emission in the quark matter core, we find that reproducing superburst properties requires a definite relation between three poorly constrained properties of strange quark matter: its thermal conductivity ($K$), its slow neutrino emissivity ($\epsilon_\nu \simeq Q_\nu \times T_9^8$) and the energy released by converting a nucleon into quark matter ($Q_\mathrm{SQM}$). The phenomenlogically required correlations are shown in Figure~\ref{Fig4}.

If strange stars exist, it is possible that either all compact stars are strange stars, or only the most massive neutron stars are strange stars, depending on whether the unknown critical density at which the transition to strange matter occurs is reached in core-collapse supernovae or whether accretion of matter is required to increase the neutron star central density to the transition point. 
Nevertheless, in both of these extreme cases, and assuming strange stars exist, the most massive compact stars must be strange stars. Reciprocally, if the most massive compact stars can be proved not to be strange stars we would be able to conclude that strange stars simply do not exist at all.
The LMXBs are good candidates for containing 
very massive compact stars, and hence possibly strange stars, 
because they are long-lived systems which undergo a significant amount of mass transfer (as much as $0.2$--$0.7$ $M_\odot$; \citealt{CST94,vdHB95}), although how much matter accretes onto the compact object is an open question. 
Two recent mass measurements of millisecond radio pulsars with white dwarf companions (likely to be descendents of LMXBs) give very different values of
 $2.1\pm 0.2$ $M_\odot$ for the 3.4 ms PSR J0751+1807 \citep{Netal05},
 confirming the formation of very massive compact stars in LMXBs,
and $1.438\pm0.024$ $M_\odot$ \citep{JHBOK05}, for the 2.95 ms period PSR J1909-3744, showing that not all LMXBs will result in very massive stars.
The high mass of PSR J0751+1807 has the very important consequence that if strange stars exist the EOS must be very stiff and, as shown in Figure~\ref{Fig1}, this implies that the central density of strange stars does not change by much more than a factor two when the mass increases from 1.1 to 2.1 $M_\odot$. In other words, all strange stars must be very similar to each other, and behave similarly independently of their mass, in contradistinction to neutron stars where various phases of exotic matter may appear at different densities, and hence different masses, and result in very distinct behaviors.
So far, we cannot claim that superbursts provide concrete evidence in favor of the existence of strange stars,
although superbursts from strange stars do offer a ``natural'' explanation for the observed low ignition column depths for carbon. More theoretical work on neutron star models for superbursts is needed before we can say definitively that a strange star solution is required.

Several interesting questions remain to be addressed in future work on accreting strange stars. Our models assume steady-state accretion, but in fact several objects are undergoing time-dependent accretion. Modeling these transient systems will give us extra, and very tight, constraints, particularly on the  star's specific heat. An important example is KS1731-260 which was observed to accrete for 13 years before transitioning into quiescence in early 2001. Its quiescent flux has been measured and is too low to explain the superburst that was observed from this source in early 2001 \citep{Retal02,B04}. In cases for which the nuclear crust is thin, $100$--$200$ meters, we find that the whole crust is liquid: a nuclear ocean, electrostatically suspended over quark matter whose density is at least four orders of magnitude higher, must have a very distinct hydrodynamic
behavior which should be investigated. Finally, finely constraining the properties of strange quark matter through modeling of these phenomena, we will be able to study the cooling of isolated strange stars. This will tell us if average compact stars, with masses between 1.1-1.4 $M_\odot$, are or are not strange stars.

%%%%%%%%%%%%%%%%%%%%%%%%%%%%%%%%%%%%%%%%
%\vspace{-0.2in}
\begin{acknowledgments} \vspace{-0.2in}
We thank Ed Brown, Tony Piro, Sanjay Reddy, and Bob Rutledge
for useful discussions. DP's work is partially supported by grants from the UNAM-DGAPA (\# IN112502) and the Mexican Conacyt (\# 27987-E). AC acknowledges support from McGill University startup funds, an NSERC Discovery Grant, Le Fonds Qu\'eb\'ecois de la Recherche sur la Nature et les Technologies, and the Canadian Institute for Advanced Research.
\end{acknowledgments}

%%%%%%%%%%%%%%%%%%%%%%%%%%%%%%%%%%%%%%%%%
 
%%%%%%%%%%%%%%%%%%%%%%%%%%%%%%%%%%%%%%%%%

%%%%%%%%%%%%%%%%%%%%%%%%%%%%%%%%%%%%%%%%%
%%%%%%%%%%%%%%%%%%%%%%%%%%%%%%%%%%%%%%%%%
\end{document}